\documentclass[conference]{IEEEtran}

\usepackage{epstopdf}
\usepackage{cite}
\usepackage{amsmath,amssymb,amsfonts,amsthm}
\usepackage{graphicx}
\usepackage{textcomp}
\usepackage{xcolor}

\def\BibTeX{{\rm B\kern-.05em{\sc i\kern-.025em b}\kern-.08em
		T\kern-.1667em\lower.7ex\hbox{E}\kern-.125emX}}
\usepackage{todonotes}
\usepackage{booktabs}
\usepackage{multirow}
\usepackage{listings}
\usepackage[inline]{enumitem}
\usepackage{adjustbox} %For two column figures
\usepackage{datetime}
\usepackage{siunitx}
\renewcommand{\dateseparator}{-}
\newcommand{\todayiso}{\the\year \dateseparator \twodigit\month \dateseparator \twodigit\day ~~ \currenttime}
\usepackage{fancyhdr}
\usepackage{algorithm}% http://ctan.org/pkg/algorithms
\usepackage{algpseudocode}% http://ctan.org/pkg/algorithmicx

\usepackage{multirow}
\usepackage{url}
\usepackage{amsthm}
\usepackage{tikz,ifthen,etoolbox,color,pgfplots} 
\usepackage{gensymb}
\usepackage{booktabs}
\usepackage{enumitem}

\allowdisplaybreaks
\algnewcommand{\LeftComment}[1]{\Statex \(\triangleright\) #1}
\let \arrow \leftrightarrow

\begin{document}
\raggedbottom

\title{A novel load distribution strategy for aggregators using IoT-enabled mobile devices \\
\thanks{This work was financially supported in part by the Singapore National Research Foundation under its Campus for Research Excellence And Technological Enterprise (CREATE) programme.}} 
%With the support of the Technische Universit{\"a}t  {\"M}unchen - Institute  for  Advanced Study, funded by the German Excellence Initiative and the European Union Seventh Framework Programme under grant agreement n\degree~291763}}

 \author{
	 \IEEEauthorblockN{Nitin Shivaraman{$^{1}$}, {Jakob Fittler$^{2}$}, {Saravanan Ramanathan$^{3}$}, {Arvind Easwaran$^{3}$}, {Sebastian Steinhorst$^{2}$}}\\	
	 \IEEEauthorblockA{$^{1}$TUMCREATE, Singapore,
	 	$^{2}$Technical University of Munich, Germany, 
		 $^{3}$Nanyang Technological University, Singapore\\
	 Email: \normalsize {nitin.shivaraman@tum-create.edu.sg, \{jakob.fittler, sebastian.steinhorst\}@tum.de, \{saravanan.r,arvinde\}@ntu.edu.sg}}
	 \vspace{-0.9cm}
 }% <- end of author section

\IEEEoverridecommandlockouts
%\IEEEpubid{\makebox[\columnwidth]{XXX-X-XXXX-XXXX-X/21/\$33.00~\copyright2021 IEEE \hfill} \hspace{\columnsep}\makebox[\columnwidth]{ }}

\maketitle

%\IEEEpubidadjcol

%\IEEEtitleabstractindextext{%
\begin{abstract}
The rapid proliferation of Internet-of-things (IoT) as well as mobile devices such as Electric Vehicles (EVs), has led to unpredictable load at the grid. The demand to supply ratio is particularly exacerbated at a few grid aggregators (charging stations) with excessive demand due to the geographic location, peak time, etc. Existing solutions on demand response cannot achieve significant improvements based only on time-shifting the loads without considering the device properties such as charging modes and movement capabilities to enable geographic migration. Additionally, the information on the spare capacity at a few aggregators can aid in re-channeling the load from other aggregators facing excess demand to allow migration of devices. In this paper, we model these flexible properties of the devices as a mixed-integer non-linear problem (MINLP) to minimize excess load and the improve the utility (benefit) across all devices. We propose an online distributed low-complexity heuristic that prioritizes devices based on demand and deadlines to minimize the cumulative loss in utility. The proposed heuristic is tested on an exhaustive set of synthetic data and compared with solutions from a solver/optimization tool for the same runtime to show the impracticality of using a solver. A real-world EV testbed data is also tested with our proposed solution and other scheduling solutions to show the practicality of generating a feasible schedule and a loss improvement of at least 57.23\%.
%  with multiple power modes across multiple aggregators is NP-hard
% . We show that the optimal scheduling problem
%Meeting energy demands requested by devices with limited impact on the grid and within a specified time is a crucial requirement for demand response.
%To this end, we designed a load scheduling algorithm on the demand side to maximize the cumulative utility (benefit) of the devices
% As a remedy to operate on constrained IoT platforms, we propose
%Extending this scheduling across multiple aggregators further increases the hardness of the problem.
%as welland show that their performance is comparable.

%increased connectivity between devices and the grid 
%has led to a bottleneck in the peak capacity at a few aggregators
%and spare aggregator capacity is explored 
\end{abstract}
\begin{IEEEkeywords}
	IoT, Load scheduling, Smart Grids, Mobility.
\end{IEEEkeywords}
%}

%We provide a bound on the performance of our proposed algorithm.
% {and the standard branch and bound solution used for knapsack problems. }

\section{Introduction}
\label{sec:intro}

% Problems due to uncertainity in EV load
The availability of various communication technologies from IoT has enabled a plethora of devices to interact with the grid for energy trading.
%With the growing adoption of IoT, Electric Vehicles (EVs) impose a high load and 
Scheduling demand requests from all the devices could result in outages and/or prohibitive costs due to exceeding the available peak capacity of the grid~\cite{6940323}. % \cite{9405374}can consume power up to 7 average North American households~\cite{iea2020}. %At a 50\% higher fuel efficiency than its gasoline counterparts
%Scheduling all requests could result in a load demand that exceeds the available peak capacity leading to outages and/or prohibitive costs. 
%The price difference between excess usage and normal usage can be up to a factor of 200 or more~\cite{tariff}.
Further, the arrival of mobile devices such as Electric Vehicles (EVs) is generally not known apriori, leading to uncertainty in demand planning. 
The penetration of renewables has alleviated this issue with distributed energy resources (DERs) that can supply additional energy in the form of aggregators (micro-and nano-grids). %~\cite{6963416}.
This bridges the disparity between the exponential rise in the number of EVs and the limited charging stations.
Traditional demand-response entails devices receiving power from a specific (geographically limited) grid location.
%These solutions aim to ensure that the peak capacity of the aggregator does not exceed while maximizing the number of devices that satisfy their demand.
These solutions aim to maximize the devices meeting their demand while keeping the demand within the aggregator's peak capacity.
However, with the availability of multiple aggregators, mobile devices (represented by EVs) communicate with the aggregators and move to different locations to improve the demand-side management (DSM)~\cite{8281479}.

\begin{figure}[ht]
	\centering  
	\includegraphics[width=1\linewidth]{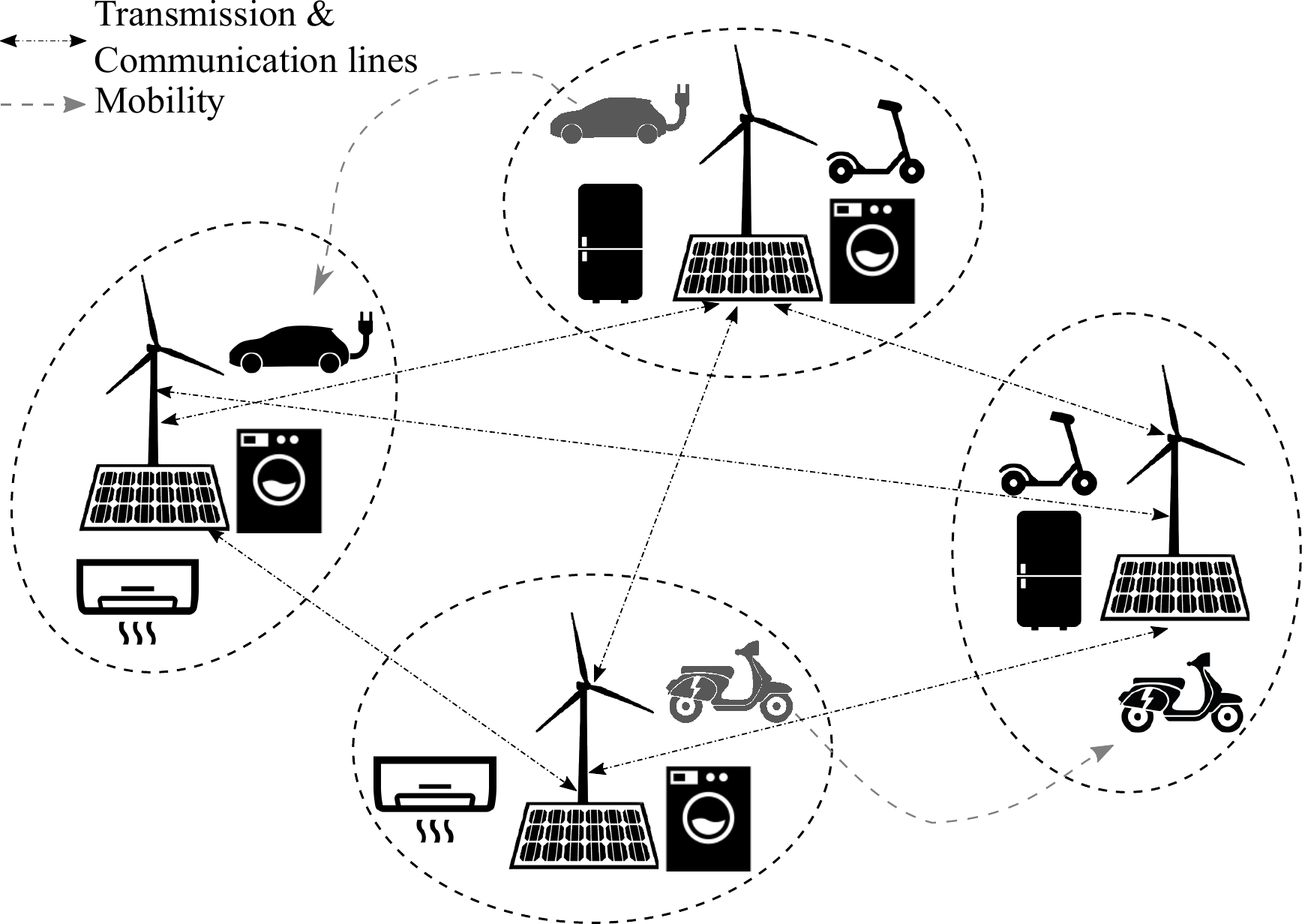}
	\caption{The grid aggregator and the devices (including mobile devices formed into clusters)}
	\label{fig:load_balancing}
	%\vspace{-0.5cm}
	\vspace{-3mm}
\end{figure}

The primary goal of DSM is to efficiently utilize the existing grid capacity to meet the demands of the devices without any grid enhancements~\cite{7063260}. %~\cite{7042318}.
The demand satisfaction is based on fulfilling the energy demand within a specified time (deadline).
Existing works~\cite{8478294} perform DSM by utilizing the day-ahead schedule such that the demand for some devices can be shifted either to an earlier or later time. 
Additionally, most of the existing solutions consider only a single mode of operation (single value of energy consumption) for the energy demand of the device during scheduling. %~\cite{8281479,7063260}

% Motivation and the problem

Most electrical devices are heterogeneous operating in multiple power modes, \emph{e.g.,} a refrigerator can operate in defrost, quick freeze, etc.%~\cite{devicemodes}. %with different capabilities 
With mobile devices, the demand can be shifted not only in time but geographically as well.
Hence, we model the device characteristics such as multiple power modes and mobility in addition to the demand and the deadlines for DSM.
The combined model is used to maximize the utility/benefit of the devices obtained by fulfilling their demanded energy within the deadline.

% Our system architecture
We consider a two-tier hierarchical system consisting of an aggregator at the higher level (supplier) and devices at the lower level (consumer) as shown in Figure~\ref{fig:load_balancing}. 
An aggregator interfaces with the grid and manages the demand from a set of devices as a cluster based on its maximum capacity.
Devices can turn on/arrive at any time and request their demand to the associated cluster's aggregator.
The power supply is realized through an electrical transmission line while the communication is either through a wired or wireless channel.

%While the demand from EVs are higher, their mobility enables load shifting to different 
%Can be equivalent to servers. arrival time is incoming request time while the departure time can be treated as the deadline.

% Goal of the paper and the architecture
In this paper, we develop an efficient scheduling algorithm to maximize the utility of the devices and aggregators while not exceeding its peak capacity at any point. 
We consider heterogeneous devices  having a different power consumption profile (\emph{e.g.,} EVs, Washing machine, etc.) and various modes of operation (\emph{e.g.,} low-power mode, sleep mode, etc.).
Based on the device parameters such as requested demand, deadline, operating modes, etc., the aggregator assigns priorities to devices and schedules them to maximize their utility.
To summarize, we make the following contributions:
%\begin{enumerate*}[label=(\roman*), itemjoin*={{. }}]
\begin{itemize}
	\item Generic device modeling: We integrate various device attributes in a model to maximize device utility (benefit).
	\item Model formulation and solution: We formulate the model and propose an online low complexity heuristic to perform the optimization.
	\item Experimental verification: We experimentally show the performance of our heuristic for runtime improvement over a solver and a utility loss improvement of over 57.23\% over standard scheduling mechanisms. % w. We also simulate the proposed solution as well as the model to show the impact of runtime and the quality of the result. 
\end{itemize}
%\begin{enumerate*}

% Organization of the paper - Do we need this?

%PECAN street dataset + Gaussian mixture model, ADRES-CONCEPT dataset.

%Paper~\cite{Karaboga2007}
\section{Related Work}
\label{sec:relatedwork}

% Works related to electric grid ---- the literature applicable to 
Load scheduling for DSM in electric grids is a well-studied problem in literature~\cite{7063260}.
The existing works in DSM can be categorized broadly based on stationary devices being grid-powered and mobile devices (such as EVs) being both grid and battery-powered. %The former analyzes the energy demand from stationary devices, while the latter addresses the demand from battery-powered mobile devices such as EVs.

Various solutions in DSM exploit load flexibility to shift the demand to a later time to minimize the peak consumption at any given time~\cite{7997120,ADIKA2014232}.
Yu and Hong~\cite{YU2016702} aim to balance the demand and supply by formulating the interaction between the aggregator and devices as a Stackelberg game where both entities maximize their utility until an equilibrium is reached.
%Noor et. al.~\cite{NOOR20181385} introduced blockchain to the game-theoretic framework for a trustless peer-to-peer consensus mechanism between aggregator and the devices.
Zhao et. al.~\cite{7997120} proposed augmenting central energy storage to the grid and virtually distributing its capacity to the devices at different time steps to maximize the operational profit.
Kim and Dvorkin~\cite{8585791} extended the concept of~\cite{7997120} with mobile energy storage that provisions for additional capacity in emergencies such as a natural disaster services to minimize the investment and operational cost.
%To overcome the computational issue of game theory-based solutions, Newton method~\cite{7406760} and proximal decomposition algorithm~\cite{6963474} were used to accelerate the run-time convergence to the Nash equilibrium, device consumption costs and to find the optimal EV schedule that reduces the gap between peak load to average load.
%Roh and Lee~\cite{7152979} categorize and model devices based on their flexibility of their preemption, scheduling and causality to maximize their utility using generalized bender's decomposition.
%Based on a similar categorization of load and to minimize peak load, a water-filling algorithm was used to provide an exact solution~\cite{7508889}.
Adika and Wang\cite{ADIKA2014232} designed a scheduling mechanism using linear programming (LP) for grid-powered and battery-storage devices to exploit charging and discharging in off-peak and peak times respectively, to minimize the consumption cost.
Chiu et.al.~\cite{7457198} not only maximize device utility but also minimize the consumption cost and carbon emissions at the grid in a multi-objective formulation and provide a distributed solution using Lagrange decomposition.
%Meta-heuristics such as genetic algorithm~\cite{OGUNJUYIGBE2017352} etc. are utilized to model and maximize device utility and obtain a schedule for load-shifting.
Although the above works minimize the aggregators' peak load, devices are restricted to only a single mode of operation and do not account for the stochastic arrival of mobile devices. % and deadlines
Practically, devices have multiple modes of operation that can be exploited to serve power to more devices within their deadlines~\cite{OZKAN2016693}. % and need to be served within a fixed time frame.
%Additionally, none of the above works consider the stochastic arrival of devices such as EVs.

% works related to EVs
%Other works in the literature focus on EV scheduling to minimize the peak load on the aggregators.
In contrast to grids where good estimates of daily, weekly or monthly schedules are available, there is more uncertainty with mobile devices such as EVs due to their stochastic arrivals and demands.
The authors of~\cite{8048016} alleviate this issue and minimize the consumption cost by letting devices estimate the day-ahead load and play a game with aggregators to minimize the deviation between actual load and estimated load.
Other methods adjust the generation of renewable resources depending on the demand from EVs~\cite{SCHULLER2015335,8478294}.
Zhang and Cai~\cite{8478294} focus on increasing the profit for aggregators by utilizing Model Predictive Control (MPC) to estimate renewable generation to adjust the EV schedule whereas, Schuller et.al.~\cite{SCHULLER2015335} focus on reducing reliance on the grid using empirical EV data and LP to maximize renewable utilization and optimal EV scheduling.
%Zheng et.al.~\cite{8307186} propose a distributed solution at the aggregator level using MPC and fuzzy logic to compute the optimal EV schedule in order to minimize the computational overhead of centralized solutions.
Similar to grid-powered devices, a Stackelberg game was used between aggregators and EVs to set prices proportional to the demand to maximize the aggregator utilization and hence, the number of EVs served~\cite{6940323}.
%However, the authors only use price-based arrival probabilities for EVs and do not track the EVs' arrival at different aggregators.
%Game theory was also used to mitigate anomalous EVs consuming the bulk of power and aim to achieve a minimum utility for any EV by committing an energy budget upon arrival and re-distributing the remaining energy budget to decide on the admittance of all the new incoming EVs~\cite{8610387}.
A priority function is introduced based on the demand and deadline of the EVs and formulated as a binary optimization problem~\cite{7496956}. %They map the continuous values of an LP to binary discrete values.
%Rassaei et. al.~\cite{7128736} consider stochastic static loads (non-mobile) as well as EVs (mobile) to minimize the cost of consumption for devices and peak load for aggregator through a decentralized game-theoretic solution and a centralized interior-point solution, respectively.
%A probabilistic threshold was used for price comparison to decide whether to provide energy to an EV for unknown pricing while using a combinatorial search method for known pricing~\cite{8288324}.
Zhu et.al.~\cite{8454434} consider multiple modes of charging linked to the time of day in a multi-stage optimization with EV scheduling where EVs change the aggregator at arrival time if the aggregator is fully loaded.
%Zhu et.al.~\cite{8454434} formulate a multi-stage optimization with EV scheduling followed by fixed device scheduling consider the possibility of changing the EV aggregator at arrival time if the aggregator is fully loaded. Although they consider multiple modes of charging, they are linked to time of day rather than instantaneous demand.
%Based on the fact that price information is not available for EVs to schedule their charging in real-time, Yi et. al.~\cite{8737558} propose deterministic online algorithms based on the charging rate and prove a bounded performance. They also provide an optimal offline algorithm when the pricing information is known.
The above works on EVs are focused on optimizations to obtain EV schedules assuming EVs cannot change their geographical location.
However, by exploiting the mobility property of the EVs, movement across different aggregators can alleviate the peak load and achieve higher utility for EVs.
Load scheduling of heterogeneous devices coupled with mobility is the novelty of our proposed work.

To the best of our knowledge, no work in the literature exploits device mobility while considering multiple modes of operation of devices.
% Please, do not modify these counters
\newcounter{arabicCounter}
\newcounter{romanCounter}
\setcounter{romanCounter}{0}

\section{Model}
\label{sec:model}

In this section, we introduce the model formulation and the objective function.
The important notations used in the paper are listed in Table~\ref{tab:Notations}.
\subsection{System}
\label{subsec:systemLevel}

%The system is mapped into clusters where each aggregator represents a cluster head of one cluster and the devices represent the cluster members associated with a cluster head.
The system architecture comprises $ J $ aggregators and $ K $ devices connected to each other as shown in Figure~\ref{fig:load_balancing}.
All the devices in the system are in the set $\mathcal{D} = \{d_1, d_2, \ldots, d_k, \ldots d_K\}$ and all aggregators are represented in the set $ \mathcal{A} = \{a_1, a_2, \ldots, a_j, \ldots a_L\}$. 
%A device $ d_k $ and aggregator $ a_l $ are represented by their unique IDs $ k $ and $ l $, respectively.
Every device in $d_k \in \mathcal{D}$ receives power from one of the aggregators $a_j \in\mathcal{A} $ by communicating information such as demand, deadline, available power at the aggregator, etc. 
$\{d_{k,j}\}$ denotes the set of all devices (including new/mobile device arrivals) in the cluster associated to $ a_j $.

Every aggregator $a_j$ has a fixed power budget $ \hat{\alpha_j} $ used for serving power demands of devices $\{d_{k,j}\}$. 
Every aggregator $ a_j $ is assumed to be connected to every other aggregator in set $ \mathcal{A} $ using a backhaul network.
%. The power budget of aggregator $ a_j $ is

% subscript for device , subscript for aggregator
% C i->j
% alpha as a set, aggregator as a set, devices as a set
% system level, parametes, device levleparameters, constants.
% i power level, j aggregator, k devices
% time is just t

  % \begin{gather}\label{eq:adamax}
    % w_{k, u} = w_{k,u-1} - \frac{\eta}{x_{u}} \hat{m}_{u}\\
    % \intertext{where,}
    % \begin{aligned}
    % & g_{k,u} = \delta_w f_{k}(w_{u-1}) \hspace{0.3cm} \textnormal{(gradient of the objective function)}\\
    % &m_{u} = \beta_1 m_{u-1} + (1-\beta_1) g_{k,u}  \hspace{0.8cm} \textnormal{(first order estimate)}\\
    % &\hat{m}_{u} = \frac{m_{u}}{1 - \beta_1^{u}}   \hspace{0.4cm} \textnormal{(bias correction for first order estimate)}\\
    % &x_u = max \bigg(\beta_2 x_{u-1},\left|g_{k,u}\right| \bigg)   \hspace{0.3cm} \textnormal{(weighted infinity norm)}\\
    % \end{aligned}\notag
    % \end{gather}

\subsection{Device level}
\label{subsec:deviceLevel}

Devices request energy with a timing constraint/deadline  $ T_k $ to fulfill their demand.
%They are heterogeneous with each device having a different power consumption profile (\emph{e.g.,} EVs, Washing machine, etc.) and various modes of operation (\emph{e.g.,} low-power mode, sleep mode, etc.).
Each device makes use of one of its power modes at every time instant to fulfill its energy demand of $ E_k $. 
Few devices have an additional property of mobility where they can move across different clusters if power unavailable at the aggregator of the incumbent cluster. 
%Each device $ d_k $ has a deadline $ T_k $ associated with its task before which it has to fulfill its energy demand. 
%Devices that have sufficient time before their deadline to receive power can be treated as flexible loads since they can receive power at a later time, increasing the elasticity on the demand side.
Hence, a device request $ \theta_k $ is represented as:

\begin{equation}\label{eq:request}
\theta_k = (R_k, T_k, m_k, I_k, E_k, \kappa_k, \mathcal{\alpha}_k)
\end{equation}

where $ R_k $ denotes the arrival time slot when the device requests power. 
This implies that either a new device switches on at $R_k$ or an existing device moves from one aggregator and arrives at another at $R_k$.
$I_k$ is the initial energy available with the device at the time of arrival (e.g. a battery's State-of-Charge).
The total energy $ E_k^{total} $ is the total energy capacity of a device which is the sum of its initial charge $ I_k $ and the demanded energy $ E_k $. 
The positive constant $ \kappa_k $ is used to implement different criticality among devices, e.g., emergency light (high $\kappa$) vs reading light (medium $\kappa$) vs washing machine (low $\kappa$). 
$\kappa_k$ also ensures that devices with higher demand do not always translate to a higher priority.
We assume that the devices demand energy that is feasible within $T_k$ with one of its power modes, i.e., $E_k \le p_k(t) \cdot (T_k - R_k)$ where $p_k(t)$ is the chosen power mode at time $t$ from the set $\alpha_{k}$, \emph{i.e.,} $p_k(t) \exists \alpha_{k,i} [\alpha_{k,i} \in \alpha{k}]$
%The arrival time is also be used to denote the time slot at which a moving device will arrive at a certain cluster. 
$m_k$ denotes if the device is mobile ($ m_k = 1 $), or non-mobile ($ m_k = 0 $).
% on-demand fixed load and $\mathcal{\alpha}$ represents the set of available power modes for $d_k$.

\begin{table}[]
	\caption{Notations and associated description used for modeling.}
	\vspace{-2mm}
	\label{tab:results}
	\resizebox{0.92\linewidth}{!}{%
		\begin{tabular}{p{3.8cm} p{5cm}}
			%\begin{minipage}[b]{1.0\linewidth}\centering
			\hline
			\centering
			Notation & Description \\
			\hline
			System parameters: & \\
			$ \mathcal{A} = \{a_{1 \ldots J}\} $					& Set of $J$ aggregators \\
			$ \hat{\alpha_j} $		& Power capacity of aggregator $ a_j $ \\
			$ c_{a_j \arrow a_{\hat{j}}} $				& Cost for movement from $ a_{j} $ to $ a_{\hat{j}} $ \\
			$ \mathcal{C} = \{c_{a_{j} \arrow a_{\hat{j}}}, \ldots\} $         & Set of all movement costs across all aggregators \\
			$ \mathcal{T} $			& Time horizon with $\tau$ slots indexed by $t$ \\
			%	$ T_0 $					& Length of one time slot -do we need it? \\
			%	$ \tau $				& Number of time slots \\
			
			Device parameters: &  \\
			$ \mathcal{D} = \{d_{1 \ldots K}\}$			 		& Set of $K$ devices \\
			%	$ \{d_{k,j}\} $				& Set of devices associated to aggregator $ a_j $	\\
			$ E_k $			    	& Energy demanded of $ d_k $ \\
			$ I_k $			    	& Initial energy of $ d_k $  \\
			$ T_k $	   			    & Deadline of $ d_k $  \\
			$ R_k $                 & Arrival/Release time of $ d_k $  \\
			$ m_k $			    	& Binary variable denoting if $ d_k $ is mobile \\
			$ \mathcal{\alpha}_k = \{\alpha_{k,0} \ldots \alpha_{k,i} \ldots \alpha{k,n}\} $			& Set of $n$ charging modes of $d_k$ \\
			$ \kappa_k $				& Criticality of $ d_k $ indicating the utility loss rate \\
			$ \gamma_{mk}[a_{j},a_{\hat{j}},t]  $             & Decision variable if device $d_k$ moves from $a_{j}$ to $a_{\hat{j}}$ \\
			$ \gamma_{pk}[i,a_j,t]  $             & Decision variable if device $d_k$ receives its $i$th power mode from aggregator $a_j$ \\
			$ \mathcal{P}_k(t) $			& Total accumulated utility up to $ t $ for $d_k$ \\
			$ p_k(t) $			& Utility achieved at $ t $ for $d_k$ \\
			
			\hline
			\label{tab:Notations}
			%\caption{Important notation used in the model formulation}
		\end{tabular}
	}
\end{table}

%Each device could have variable modes of operation where it consumes different power depending on the mode of operation. 
Each device $ d_k $ has a power mode $ \alpha_{k,0} = 0 $ when the device is not served any power, as well as up to $ n \in \mathcal{N} $ power modes in increasing order of consumption, i.e., $ \alpha_{k,0} < \alpha_{k,1} < \ldots < \alpha_{k,n} $.
The number of power modes and the values for each mode are specific to each device. 
Hence, the set of all power modes of device $ d_k $ is given as $ \mathcal{\alpha}_k=\{\alpha_{k,0}, \alpha_{k,1},\ldots,\alpha_{k,n} \} $. 

%The set of all $ n $ requests of all devices $ d_k \in D $ is given as $ \Theta = \{\theta_k\},|\Theta|=n $.

\paragraph{Mobility}
Mobile devices can move across different clusters as they are equipped with energy storage such as batteries that facilitate energy required for movement. 
Any device $ d_k $ that moves from its current cluster with the aggregator $ a_{j}$ to another (target) cluster with the aggregator $ a_{\hat{j}} $ has a delay $ \delta_{a_{j} \arrow a_{\hat{j}}} $ and an associated cost $ c_{a_{j} \arrow a_{\hat{j}}} $ per unit delay to pay for the movement represented by the tuple $<\delta_{a_{j} \arrow a_{\hat{j}}},c_{a_{j} \arrow a_{\hat{j}}}>$.
The delay $\delta_{a_{j} \arrow a_{\hat{j}}}$ from current time $t_c$ is the arrival time of $d_k$ at aggregator $a_{\hat{j}}$, i.e., $R_k$ of $d_k$ at $a_{\hat{j}}$ is $t_c + \delta_{a_{j} \arrow a_{\hat{j}}}$.
Movement delays are representative of the distance of the aggregators from each other and the movement costs are proportional to the distance. 
E.g., an EV that moves longer distance incurs higher cost and delay and vice-versa.
The movement option for any mobile device to move between aggregators $j1$ and $j2$ are given by,
\begin{equation}\label{eq:movementTimeCostTuple}
\mathcal{C}_{a_{j} \arrow a_{\hat{j}}} = <\delta_{a_{j} \arrow a_{\hat{j}}},c_{a_{j} \arrow a_{\hat{j}}}>
%\mathcal{C}_{j1 \arrow j2} = \{ (\delta_{1,j1 \arrow j2},ct_{1,j1 \arrow j2}),\ldots, (\delta_{H,j1 \arrow j2},ct_{H,j1 \arrow j2}) \}
\end{equation}
The total cost of a movement is given by the product, i.e., $c_{a_{j} \arrow a_{\hat{j}}}(t) = \delta_{a_{j} \arrow a_{\hat{j}}} \cdot c_{a_{j} \arrow a_{\hat{j}}} $.
%$ h $ denotes the $ h $-th movement choice to the $ h $-th aggregator among set of all movement options $\mathcal{H}$ of the device. % i.e., different ways a device can move (e.g. high velocity, shortest path, etc.).
%Depending on the movement option chosen, the cost and the time vary.

Without loss of generality, movement times are ordered in increasing order of movement time. 
Since the above tuple specifies the movement options only between clusters with aggregators $a_j$ and $a_{\hat{j}}$, a matrix $ C $ is defined to include all movement options across all clusters and is given by:

\begin{equation}\label{eq:movementTimeCostMatrix}
\mathcal{C} = 
\begin{bmatrix}
0        &\ldots   &(1 \arrow \hat{j})\\  % &(1 \arrow 2)_{\{h\}}
(2 \arrow 1) & 0        &\vdots\\ % &\ddots
\vdots  &\ddots        &(j-1 \arrow \hat{j})\\ %&0 
(j \arrow 1) &\ldots  &0 % &(i \arrow j-1)_{\{h\}}
\end{bmatrix}
\end{equation}
%\begin{equation*}\label{eq:movementTimeCostMatrix}
%[M]_{i,j} = \mathfrak{T}_{i,j} \times C_{i,j}
%\end{equation*}

%The number of available movement options and hence, the costs, may differ across different clusters ($ a_{j1}, a_{j2} $ pairs).
%The above matrix has the same value for a given pair and hence, can be simplified into an upper triangular matrix.
Mobile devices are limited by their initial/available energy while choosing the movement options, i.e., the movement cost cannot be greater than their available energy.
The chosen movement cost is an additional energy derived from the aggregator along with its demand $E_k$. 
%It is important to notice that serving stationary devices the requested energy $ E_k^d $ leads to the completion of the task of this stationary device. 

\subsection{Utility function}

\label{subsec:utilityFunction}
Assuming a discrete-time system, the total time $ \mathcal{T} $ is divided into $ \tau $ time slots of length $ T_0 $ and denoted as $ \mathcal{T} = \{1,\ldots, \tau\} $. 
Each device achieves a certain utility (benefit) when it is served power at any time slot. 
A utility $ p_k(t) $ is achieved by a device at any given time slot $t$ for a duration $T_0$ when it is served with power $\alpha_{k,i}$ that is among the power modes $\alpha_k$.
The utility $ p_k(t) $ of each time slot is appended to the set of accumulated utility $\{p_k(0),p_k(1),\ldots,p_k(t)\}$.
Hence, at the current time slot $ t_c $, $\mathcal{P}_k(t_c)$ is the sum over the utility $ p_k(t) $ of all the previous time steps leading to $t_c$.

\begin{equation}\label{eq:accumulatedProgress}
\mathcal{P}_k(t_c) = \sum_{t=0}^{t_c} p_k(t) \gamma_{pk}[i, a_j, t]
\end{equation}

where $\gamma_{pk}[i, a_j, t]$ indicates that $i$th power mode was served at cluster with aggregator $a_j$ at time $t$.
In an ideal case, all devices are scheduled before their $T_k$ and served with the corresponding $E_k$, yielding a utility of $E_k$ without any loss.
However, in practice, due to congestion and devices with higher demand, scheduling all devices without deadline misses or movement is not feasible.
Devices incur a utility loss when it moves (as described earlier) or misses its deadline that is time-dependent.
%independent of the time. However, it also
The total utility loss function for devices per time slot $t$ is given by:

\begin{equation}\label{eq:utility}
\begin{split}
u_k(t) &= \beta_k^d(t) + 2 \cdot \beta_k^m(t) \\
		 & + (1-m_k)(1-\delta_{a_j \arrow a_{\hat{j}}}(t))\beta^{max} \\  %p_k(t) \cdot \gamma_{pk} \cdot T_0 +
		 % \dfrac{c_{h,i,j}(t)}{E_k} \gamma_k(t) +
		 % + c_{h,j1 \arrow j2}(t) \cdot \gamma_k(t)
		 % \dfrac{}{E_k}
\end{split}
\end{equation}

The utility loss function in Equation~\ref{eq:utility} consists of three terms: loss due to mobility, loss for missing the deadline and penalty for moving stationary devices across clusters. %Utility gain based on power consumption,

\subsubsection{Utility loss due to deadline violation}
The first term in Equation~\ref{eq:utility} incorporates the penalty for missing the deadline. It is given as

\begin{equation}\label{eq:negUtilityMissingDeadline}
\beta_k^d(t) = \begin{cases}
f(t), &  (t > T_k) \wedge (\mathcal{P}_k(t) \le E_k) \\
0	  & \text{otherwise}
\end{cases} 
\end{equation} % \forall d_k
where $f(t)$ is given by $(\mathcal{P}_k(t_c) - E_k) \cdot [\exp(\kappa_k(t_c-T_k))]$. %-\exp(\kappa((t_c-1)-T_k^d)) 
$ \mathcal{P}_k(t) $ is the accumulated utility up to the current time slot $t_c$ as defined in Equation~\ref{eq:accumulatedProgress}.
The expression in Equation~\ref{eq:negUtilityMissingDeadline} is the loss factor per time slot after the deadline is missed. % equivalent to $\alpha_{k,n}$. %and $\beta$ is a negative constant
$\kappa_k$ provides a measure of criticality in devices to signify how imperative it is to serve a device $d_k$ to minimize the significant losses in utility.
A higher value for $ \kappa_k $ translates to a higher rate of loss for exceeding the deadline. 

\subsubsection{Permanent utility loss due to mobility}
Mobile devices consume additional energy than their initial demand to finish their tasks owing to energy loss during movement. 
The loss in energy reduces the utility and consequently, increases the demand of the device from the aggregator.
This additional energy supplied by the aggregator due to mobility on top of the initial energy demand of all the devices leads to a utility loss on the grid which cannot be compensated. 
Hence, there is a factor of two for the second term for utility loss due to mobility.
The loss due to mobility is given as:

\begin{equation}\label{eq:permanentUtilityLossDueToMobility}
\beta_k^m(t) = c_{a_j \arrow a_{\hat{j}}} \cdot \gamma_{mk}[a_j, a_{\hat{j}}, t] %{\hat{\alpha}_{avg}}b_k^m %\hat{U} %{\beta_k^{min} 
\end{equation}

where, $c_{a_j \arrow a_{\hat{j}}}(t)$ is the cost per time slot derived from Equation~\ref{eq:movementTimeCostMatrix} and $\gamma_{mk} [a_j, a_{\hat{j}}, t]$ represents the binary variable that indicates if a device is chosen to move (= 1) or not (= 0) between clusters with aggregators $a_j$ and $a_{\hat{j}}$ at time $t$.
It is important to note that if a mobile device with an initial charge moves to another cluster before consuming power, it results in negative utility that is capped at $-I_k$.

%with $ \hat{\alpha_{avg}} $ the average budget of all clusters $ a_l \in A $ which is the ratio of the sum of all aggregator budgets over the number of aggregators. 
%$\hat{U}$ denotes the minimum achievable utility $ U_k^g $ for all $ d_k \in D $.
% Add the reason for using these terms i.e., alpha_avg and U_hat.....

\subsubsection{Utility loss for moving stationary devices}
The third term in Equation~\ref{eq:utility} prevents the stationary devices to move from its parent cluster by imposing a very high penalty on those devices. If a device $d_k$ belongs to a cluster $a_j$ and it moves to a different cluster $a_{\hat{j}}$, then the penalty is given by

\begin{equation}\label{eq:utilityLossMovingStationaryDevice}
(1-m)(1-\delta_{a_j \arrow a_{\hat{j}}})\beta^{max}
\end{equation}
with $ \delta_{a_j \arrow a_{\hat{j}}} $ is $0$ if $ (a_j = a_{\hat{j}}) $, and $ 1 $ otherwise. $ \beta^{max} $ is the maximum penalty constant configured to a very high value.

%\subsection{Assumptions}
%\label{subsec:assumptions}
%We make the following assumptions for formulating our model and objective:
%\begin{itemize}
%	\item Devices are  heterogeneous, i.e., they are of different types such as heating, lighting, etc.
%	\item The system is reliable and trustworthy, i.e., no loss of messages or tampered data.
%	\item Each device is connected only to one aggregator at any time
%	\item For every device $ d_k $ every available power mode (except $ \alpha_k^0 = 0 $) and the total requested energy are positive integer multiples of $ \alpha_k^1 $, i.e., $ \forall \alpha_k^z \in \alpha_k \setminus \{0\}, E_k^d+E_k^i, n \in \mathbb{N}: \alpha_k^z = n \alpha_k^1, E_k^T = n\alpha_k^1 $
%	\item The devices $d_k^i$ of cluster $i$ can directly communicate to other devices within the same cluster but cannot communicate to devices $d_k^j$ of a different cluster $j$.
%	\item Serving a portion of the total energy $ E_k^d $  of a device increases the $\psi_k$ by the same proportion since it is normalized by $E_k^T$. 
%\end{itemize}

\subsection{Objective function and constraints}
\label{subsec:objectiveFunctionAndConstraints}

Given all devices $ d_k \in \mathcal{D} $ and the utility function in Equation~\ref{eq:utility}, the objective is to minimize the cumulative utility loss across all devices over all the time slots in the system.
It is important to note that the objective is to allow a device with high criticality to consume power at the earliest to reduce the penalty in utility rather than maximizing the devices to meet their deadlines. 
Hence, there could be a higher number of deadline violations while minimizing the utility loss. 

Since the progress $ \mathcal{P}_k(t) $ remains the same independent of the mobility or the time of receiving power, we can translate the maximization of utility into a minimization of utility loss.
Assuming that the stationary devices do not move, the utility loss term is given by the combination of terms one and two in~\ref{eq:utility}, i.e., 
$u_{k}(t) = 2 \cdot \beta_k^m(t) + \beta_k^d(t)$.  % \rho \cdot   (1 - \rho) \cdot
The decision variables are the choice of each device $d_k$ to receive power ($\gamma_{pk}$) and the choice to move to another cluster ($\gamma_{mk}$).
%$\rho$ indicates the weight parameter to signify the precedence of the loss minimization between deadline and mobility.

The objective function and the constraints are given as:
\begin{equation}\label{eq:objectiveFunction} 
	\min \sum_{t \in \mathcal{T}} \sum_{d_k \in D} u_{k}(t) 	
\end{equation}

\setcounter{arabicCounter}{\value{equation}}	% Save most recently used arabic counter
\setcounter{equation}{\value{romanCounter}}		% Set equation counter to most recently used roman counter
\renewcommand{\theequation}{\roman{equation}}	% Change numbering to roman

%\begin{align}
%	\textrm{s.t.} \quad	& \sum_{d_k \in d_{k,j}} (1 - m) \cdot p_k \cdot \gamma_{pk}(t) + m \cdot p_k \cdot \gamma_{pk}(t) \cdot \gamma_{mk}(t) \le \hat{\alpha_j}, & \forall t, \forall a_j \in A
%	\label{constr.capacityConstraint} \\
%	& \sum_{a_j \in A}\sum_{d_k \in d_{k,j}} \gamma_{pk}(t) \le 1, & \forall t
%	\label{constr.powerProfileConstraint} \\
%%	& c_{h,i,j}(t_c) \in \mathcal{C}, & \forall t, \forall \{a_i,a_j\} \in A
%%	\label{constr.mobilityCostConstraint} \\
%%	& -\dfrac{E_k^i}{E_k^d+E_k^i} \le \mathcal{P}_k(t) \le \dfrac{E_k^d}{E_k^d+E_k^i}, & \forall t, \forall d_k \in D
%    & \sum_{t \in T} (1 - m) \cdot p_k \cdot \gamma_{pk}(t) + m \cdot p_k \cdot \gamma_{pk}(t) \cdot \gamma_{mk}(t) \le E_k,& \forall d_k \in D   
%	\label{constr.consumptionConstraint} \\
%	%& ((p(t_c)) \cdot (c_{h,i,j}(t_c)) = 0, & \forall t, \forall d_k \in D
%	& \gamma_{pk} \cdot \gamma_{mk} = 0, & \forall t, \forall d_k \in D
%	\label{constr.consumptionConstraintDuringMobility} \\
%%	& d_k^i \cap d_k^j = \emptyset, & \forall \{a_i, a_j\} \in A 
%%	\label{constr.connectivityConstraint} \\
%	& \gamma_{mk}, \gamma_{pk} \in \{ 0,1 \}, & \forall d_k \in D
%	\label{constr.mobilityStatusConstraint}
%\end{align}

\begin{align}
	\textrm{s.t.} & \sum_{j \in A} \sum_{i \in \alpha_k} \gamma_{pk}[i,a_j,t] \leq 1, \qquad \forall t, ~\forall d_k \in \mathcal{D} \label{constr.1} \\
	& \sum_{a_j \in A} \sum_{a_{\hat{j}} \in \mathcal{A}} \gamma_{mk}[a_j,a_{\hat{j}},k] = 1, \qquad \forall t, ~\forall d_k \in \mathcal{D} \label{constr.2} \\
    & \sum_{i \in \alpha_k} ((1 - m_k) \cdot p_k(t) \cdot \gamma_{pk}[i,a_j,t] \nonumber\\
    & \quad + m_k \cdot p_k(t) \cdot \gamma_{pk}[i,a_j,t] \cdot \gamma_{mk}[a_j,a_{\hat{j}},t]) \leq \hat{\alpha_j},\nonumber\\ 
    & \qquad \forall t, ~\forall \{a_j, a_{\hat{j}}\} \in \mathcal{A}, ~\forall d_k \in \mathcal{D}	\label{constr.3} \\
	& \gamma_{mk}[a_j,a_{\hat{j}},t] + \gamma_{mk}[a_{\hat{j}},a_{\hat{j}},t] \nonumber\\
	& \quad - \gamma_{mk}[a_j,a_{\hat{j}},t-1] \ge 0, \nonumber\\
	& \qquad\forall d_k \in \mathcal{D}, ~\forall \{a_j, a_{\hat{j}}\} \in \mathcal{A}, ~\forall t, ~\forall a_j != a_{\hat{j}} \label{constr.4} \\
	& \gamma_{mk}[a_j,a_{\hat{j}},t] + \sum_{a_{\hat{j}} \in \mathcal{A}}\gamma_{mk}[a_j,a_{\hat{j}},t] \nonumber\\
	& \quad - \gamma_{mk}[a_j,a_j,t-1] \geq 0, \nonumber\\
	& \qquad \forall d_k \in \mathcal{D}, ~\forall \{a_j, a_{\hat{j}}\} \in \mathcal{A}, ~\forall t, ~\forall a_j != a_{\hat{j}} \label{constr.5} \\
	& \gamma_{mk}[a_j,a_{\hat{j}},t] \cdot \delta_{a_j \arrow a_{\hat{j}}} + \bigg(\gamma_{mk}[a_j,a_{\hat{j}},t-1] * \nonumber\\
	& \quad \big(\sum_{t_m = 1}^{\delta_{a_j \arrow a_{\hat{j}}}} (\gamma_{m,k}[a_j,a_{\hat{j}},t-t_0]) - \delta_{a_j \arrow a_{\hat{j}}}\big)\bigg) \geq 0,\nonumber\\ 
	& \qquad \forall d_k \in \mathcal{D}, ~\forall \{a_j, a_{\hat{j}}\} \in \mathcal{A}, ~\forall t, ~\forall a_j != a_{\hat{j}} \label{constr.6} \\
	& \sum_{t_m = 1}^{\delta_{a_j \arrow a_{\hat{j}}}} \gamma_{mk}[a_j,a_{\hat{j}},t-t_0] \leq 0,\nonumber\\
	& \quad \forall d_k \in \mathcal{D}, ~\forall \{a_j, a_{\hat{j}}\} \in \mathcal{A}, ~\forall t, ~\forall a_j != a_{\hat{j}} \label{constr.7} \\
	& \sum_{t \in T} (1 - m_k) \cdot p_k(t) \cdot \gamma_{pk}[i,a_j,t] + \nonumber\\
	& \quad m_k \cdot p_k(t) \cdot \gamma_{pk}[i,a_j,t] \cdot \gamma_{mk}[a_j,a_{\hat{j}},t] \le E_k, \forall d_k \in \mathcal{D}  \label{constr.8}
	\vspace{-3mm}
\end{align}

\setcounter{romanCounter}{\value{equation}}		% Keeping track of last roman numbering
\setcounter{equation}{\thearabicCounter}		% Restore proper number for arabic numbering
\renewcommand{\theequation}{\arabic{equation}}	% Change numbering to arabic

where Constraint~(\ref{constr.1}) ensures that the power mode chosen in any time slot is from the set of available power modes at only one of the aggregators.
The state of a device can either be stationary or in motion as indicated by Constraint~(\ref{constr.2}).
Constraint~(\ref{constr.3}) constrains the power used by all devices of a cluster to not exceed the maximum power budget $ \hat{\alpha_j} $ of its aggregator $ a_j $ for all time slots.
Constraints~(\ref{constr.4}) and (\ref{constr.5}) shows the behavioral constraint on the state of the device in the current time slot based on the state in the previous time slot, \emph{i.e,} a device has a mutually exclusive state of being stationary or moving.
The movement of a device for at least $\delta_{a_j \arrow a_{\hat{j}}}$ between clusters with aggregators $a_j$ and $a_{\hat{j}}$ is specified as a constraint in Constraints~(\ref{constr.6}) and (\ref{constr.7}).
%Equation~(\ref{constr.powerProfileConstraint}) and Equation~(\ref{constr.mobilityCostConstraint}) ensure that used power mode and cost for movement are taken from the set of available power and movement options respectively. 
%This constraint shows all devices have progress up to 100\%. %shown by the inequality on the right while ensuring mobile devices can only use a maximum of its initial energy to move to other clusters before receiving energy from the aggregator given by the left inequality. 
%Equation~(\ref{constr.consumptionConstraintDuringMobility}) captures that a device cannot consume power during movement. 
%Equation~(\ref{constr.connectivityConstraint}) ensures that each device is associated to at most one cluster/aggregator at any time. 
%Equation~(\ref{constr.mobilityStatusConstraint}) indicates that the total energy consumption of a device does not exceed its demand. 
The total accumulated energy of a device across all time slots is bounded by $E_k$ as shown by Constraint~(\ref{constr.8}).
%All the notations used are listed in Table~\ref{tab:Notations}.

The objective function in Equation~\ref{eq:objectiveFunction} has integer constraints with the choice of power modes of a device and is formulated for a discrete-time. %~\ref{constr.mobilityCostConstraint}
Additionally, the presence of binary decision variables $\gamma_{mk}$ and $\gamma_{pk}$ makes the problem non-linear.
Hence, the problem is a mixed-integer non-linear programming problem (MINLP).
For this problem, we try to serve the device with the best utility within the constrained power budget of the aggregator.
The optimization in Equation~\ref{eq:objectiveFunction} with constraints is an NP-hard problem in the form of multiple-choice multiple-knapsack problem due to the combinatorial nature of picking one of many power modes and choosing one of many aggregators. 

\section{Low-complexity heuristic solution}

The complexity of the problem scales with the number of devices and aggregators to obtain an optimal solution.
%Due to the nature of the problem, the complexity sobtaining an optimal solution leads to high complexity that scales exponentially with the number of devices and aggregators. 
To overcome this issue, we propose an online distributed low-complexity algorithm that scales well at an aggregator level.
%Since each aggregator operates independently, the complexity reduces to $\mathcal{O}(n)$ where $n$ is the number of aggregators.

%Aggregators are responsible to handle the scheduling of devices to minimize the utility loss.
Devices submit their requests to their respective cluster's aggregator in the form of Equation~\ref{eq:request} and aggregators schedule these requests to minimize the utility loss.
Each aggregator uses a \emph{priority} function to establish the order in which the devices can receive power in the next time slot. 
Additionally, the power mode for each device is also decided for the time slot and the schedule gets disseminated to all devices within the same cluster indicating the order in which the devices receive power.
%The schedule is disseminated to devices within the same cluster.
Using this information, each device computes its corresponding utility loss.

Furthermore, mobile devices that cannot fulfill their demand within the deadline submit their requests to move to another aggregator.
Since the aggregators are connected using a backhaul network, each aggregator can compute the cost and the associated time of movement to other aggregators. 
The backhaul network consists of electrical connections to the grid as well as network connections with the grid operator to facilitate load data analysis.
Hence, using the cost and computed utility loss for movement provided by the aggregator, mobile devices are able to take independent decisions on making a move to a different aggregator.
It is important to note that aggregators do not have any information on the future arrivals of devices and computes a schedule only based on the information in the current time slot.

\paragraph{Priority function}

At every time slot, devices get assigned a priority order from the aggregator using a priority function that uses the remaining time to deadline and remaining power to fulfill the demand.
Aggregator computes the utility loss for each device at a given time slot and orders them in decreasing order of loss, i.e., the device with the highest loss gets the highest priority.
%Since the metric for priority can influence in the incurred loss in utility, we propose three different priority functions based on the demand finish time of the device.
The priority function $pr_k$ is defined for all devices with $\mathcal{P}_k(t_c) \le E_T$ as:

\begin{equation}\label{eq:priorityfunction}
pr_k = \begin{cases}
\dfrac{(E_k - \mathcal{P}_k(t_c))}{E_k} \cdot (t_c - T_k), &  (t_c > T_k) \\
\dfrac{E_k - \mathcal{P}_k(t_c)}{E_k} \cdot \dfrac{1}{(T_k - t_c)}, & (t_c < T_k) \\
\dfrac{E_k - \mathcal{P}_k(t_c)}{E_k}, & (t_c = T_k)
\end{cases} 
\end{equation}

%\begin{equation}\label{eq:priorityCalculationFinishingTime}
%pr(d_k) = \dfrac{\sum_{t=t_c}^{t_F} \beta_k^d(t)}{p_k^{max}}
%\end{equation}

%where $t_c$ is the current time slot when the priority is computed.
%Equation 
With $t_c$ significantly lower than $T_k$, devices can sustain waiting without incurring utility loss. 
Alternatively, when $t_c$ is greater than $T_k$, $d_k$ has to be served power at the earliest to prevent utility loss.
With $t_c$ at $T_k$, the priority function is the ratio of the remaining demand to complete over the total demand of the device. 
The priority function is a function of the remaining power and the remaining time to the deadline. 
Intuitively, this translates to the penalty from $\beta_m$ and $\beta_d$ which yield higher losses due to an increase of demand due to movement and deadline misses, respectively. 
In case two device $pr_k$, the tie is broken using other parameters such as $\kappa_k$ and feasible $\alpha_k$ based on the available aggregator supply.

%, $t_F$ is the estimated finish time based on the chosen priority option, $\beta_k^d(t)$ is the utility loss due to deadline at any time slot $t$ and $p_k^{max}$ is the highest power mode of the device $d_k$.
%The finish time is estimated based on three priority options: earliest finish time, latest finish time, and mixed-case finish time.

%The latest finish time assumes that the devices are provided power sequentially with every device served with its highest power mode.
%It is computed as the ratio of remaining energy demand to be fulfilled over the highest power mode of the device and is given as:
%\begin{equation}\label{eq:latestFinishingTime}
%t_{F}^{latest} = \sum_{d_k \in d_k^j} \ceil[\bigg]{ \dfrac{(\mathcal{P}_k(t) - E_T)}{p_k^{max}} }
%\end{equation}
%
%In contrast to the latest finish time, the earliest finish time assumes the devices are provided power to maximize the utilization of the aggregator power budget.
%Hence, it is computed as the ratio of sum of all the remaining energy demand of the devices over the aggregator's power budget, given as:
%\begin{equation}\label{eq:earliestFinishingTime}
%t_{F}^{earliest} = \ceil[\Bigg]{\dfrac{\sum_{d_k \in d_k^j} (\mathcal{P}_k(t) - E_T)}{\hat{\alpha}_j} }
%\end{equation}
%
%The mixed-case finish time is a weighted sum of both the latest and earliest finish time of the devices.
%With x as the weight parameter, the mixed-case finish time is computed as:
%\begin{equation}\label{eq:mixedCaseFinishingTime}
%t_f^{M} = x \cdot t_{F}^{earliest} + (1-x) \cdot t_{F}^{latest}
%\end{equation}

\begin{algorithm}[t!]
	\scalebox{0.85}{
		\begin{minipage}{0.98\columnwidth}
			\caption{Distributed Algorithm to minimize utility loss of devices (static and mobile) with multiple power modes}
			\label{alg.heuristic}
			\begin{algorithmic}[1]
				\LeftComment {System Level:}
				\For{$ a_j \in \mathcal{A}$}
					\State Initialize status\_list $\gets []$
					\State Rcv($\theta_k$) $\forall d_k in a_j$
					\State Initialize $ d_k^j$
					\State Initialize $pr\_list_j \gets []$
				\EndFor
				\LeftComment {Aggregator Level:}				
				\For{$ d_k \in \mathcal{D}$}
					\State Compute $\mathcal{P}_k(t_c)$
					\State Compute $pr_k$ $ \gets$ priority($\mathcal{P}_k(t_c), E_k, t_c, T_k$)
					\State Sort\_descending ($pr\_list_j$($pr_k$))
				\EndFor
				\State $p_k(t_c) \gets \alpha_{k,i}\forall d_k \in pr\_list$

				\If{$(\hat{\alpha_j}$ - $\sum_{d_k \in d_k^j} p_k(t_c)) > 0$}
						\State $p_k(t_c) \gets \alpha_{k,i+1}$ $\forall d_k in pr\_list$
				\EndIf
				\State status\_list $\gets$ get\_agg\_status($\mathcal{A} \setminus a_j$)
				\State Send (status\_list) $\forall d_k^j$
				
				\LeftComment{Device Level:}
				\State $c_{a_{j} \arrow a_{\hat{j}}}, \hat{\alpha_{\hat{j}}} \gets$ Rcv(status\_list)
				\If{$(\gamma_{pk} == 0) \&\& (m_k == 1)$}
					\For {$c \in \mathcal{C}$}
						\State Compute $\beta_k^m(t_c) \gets$ calc\_move($c_{a_{j} \arrow a_{\hat{j}}}$,$\hat{\alpha_{\hat{j}}}$)
						\State Compute $\beta_k^d(t_c)$
					\EndFor
					\If{($\beta_k^d(t_c) > \beta_k^m(t_c)$)}
						\State Send($\theta_k$)
					\EndIf
				\EndIf
			\end{algorithmic}
	\end{minipage}}
\end{algorithm}

The distributed low-complexity algorithm that is executed at every time slot is shown in Algorithm~\ref{alg.heuristic}.
At a system level (lines 1-6), the aggregators (re-)initialize the devices in its cluster-based on device arrival requests and the associated priority list. 
They also initialize the \emph{status\_list} variable to an empty set that is later populated with aggregator utilizations (load).
Each aggregator computes the priority for each of the devices within its cluster-based on Equation~\ref{eq:priorityfunction} and sorts them in descending order.
After computing the remaining time and power to meet the device demand, the parameters are passed to the priority function shown in Equation~\ref{eq:priorityfunction}.
Devices in the list are assigned power with the lowest of their feasible power modes since few modes may not be feasible based on the progress. %, \emph{i.e.,} starting from the highest priority with the least feasible power mode.
Subsequently, any available power at the aggregator is used to upgrade to the higher power mode for devices in the same order of priority.
Lastly, the status\_list is updated by communicating the utilization among the aggregators and shared with the devices for mobility.
Aggregator steps are illustrated from lines 7-17 of Algorithm~\ref{alg.heuristic}.
At the device level (lines 18-24), mobile devices that did not receive power use the information from the status\_list to compute the loss due to movement.
If the loss can be minimized at a neighboring aggregator, it sends a request to join the neighboring cluster.
The above steps are common to all devices independent of their mobility capabilities shown in steps 1-11 in Algorithm~\ref{alg.heuristic}.

The time complexity of the heuristic is $\mathcal{O}(n\log{}n)$.
$\log{}n$ is the complexity of the sorting algorithm used for prioritizing devices while repeating it for $n$ devices. 

%For mobile devices, if a movement has to be made, the utility loss to other aggregators are computed and transmitted to the device.
%Based on the incurred utility loss to move across different aggregators, time to move and available charge, the device takes the feasible movement that minimizes the loss as shown in lines 12-20 of Algorithm~\ref{alg.heuristic}.
\section{Experiments}
\paragraph{Experimental Setup}
In this section, we evaluate our heuristic solution in minimizing the utility loss with a synthetic and a real-world dataset. 
To compare the performance of the proposed algorithm, we also solve the optimization problem in Gurobi solver as a baseline~\cite{gurobi}.
Python 3.7 was used for both Gurobi and the proposed solution. 
Firstly, synthetic data was generated using the DRS Algorithm~\cite{9355491} based on the parameters defined in Table~\ref{tab.simparams}.
The parameters used in Table~\ref{tab.simparams} are derived from the real-world datasets to mimic devices such as HVACs~\cite{hvac} and EVs~\cite{evdatabase}.
%which generates a uniform distribution of utilization for the aggregators and devices. Synthetic input data was used 

Each cluster load was split into three classes: Lightly loaded, medium loaded and heavily loaded with utilizations shown in Table~\ref{tab.simparams}. % aggregator utilitization (load)
This indicates that the devices use up the corresponding utilization of the aggregator capacity, \emph{i.e.,} for 0.8 utilization and 500kWh as aggregator capacity, devices consume 80\% (400kWh) of the capacity throughout the time horizon ($\mathcal{T}$).
Increasing the utilization to the higher end (H) on all aggregators leads to a non-feasible schedule since each device would have a high demand, leading to some devices (low-criticality) never getting scheduled. 
Devices are generated to achieve utilization by taking into consideration the maximum consumption capacity of various device classes (HVAC, EV, etc.).
For example, devices such as iron consume 1kWh, airconditioners consume about 3kWh and EVs can consume up to 50kWh.
A deadline is randomly chosen to create a periodic demand from the devices (arrival = deadline) and corresponding power modes are assigned to ensure that demand can be met.
Mobility costs and movement time between the clusters are inversely related to distance \emph{i.e.,} the farthest cluster incurs the highest mobility cost and time for movement (4, 0.15) and vice-versa (1, 0.15).

\begin{table}[t]
	\begin{minipage}[b]{1.0\linewidth}\centering
		\begin{center}
			\caption{Parameters used in our experimental setup for evaluating the proposed algorithm.}
			\vspace{-5mm}
			\label{tab.simparams}
			\scalebox{0.85}{%
				\begin{tabular}{@{}ll@{}}
					\toprule
					\textbf{Parameter}          & \textbf{Value} \\ \hline
					$ \mathcal{A} $           & 5         \\ 
					$ \hat{\alpha}_l $      & $500k \mathrm{Wh}$  \\ 
					No. of Devices        &  \{20, 40, 60, 80, 100\}  \\
					No. of Timeslots ($\tau$)      &  50 \\
					Deadlines/Periodicity  & \{6,12,24,48\}  \\
					Classes of aggregator load     &  \{[0.5-1] (L), [1-1.25] (M), [1.25-1.5] (H)\}   \\ 
					Aggregator load combinations & {[L,L,L,M,H],[L,L,M,M,H],[L,M,M,M,H],[L,L,M,H,H]} \\
					$ T_0 $           & $ 0.5\mathrm{h} $   \\ 
					$ (\delta_{a_{j} \arrow a_{\hat{j}}}, t_{a_j \arrow a_{hat{j}}}) $    &  \{(0,0), (1,0.15), (2,0.15), (3,0.15), (4,0.15)\} (slots, $\mathrm{kWh}$)   \\
					$ \alpha^{min} $ & $ 1\mathrm{kW} $  \\
					$ \{|\alpha_k\} $  &  \{1,2,3,5,10,20,50\}$\mathrm{kW}$  \\ 
					Proportion of Mobile Devices       & \{$ 25\%, 50\%, 75\%, 100\% $\} \\
					$ \kappa_k $    &  \{1.6, 1.8, 2.0\}  \\ \hline
				\end{tabular}
			}
		\end{center}
	\end{minipage}
\end{table}

\subsection{Heuristic Performance}
%To test the performance of the heuristic in an unbiased way, an exhaustive synthetic dataset was generated using the DRS (Dirichlet rescale) algorithm~\cite{9355491}.
%DRS is based on the popular Uunifast algorithm~\cite{Bini2005} that has been widely used for task set generation with uniform distribution for a single resource.
%Since multiple aggregators constitute multiple resources, DRS is more suitable. 

\begin{figure}[ht]
	\centering  
	\includegraphics[width=1.0\linewidth]{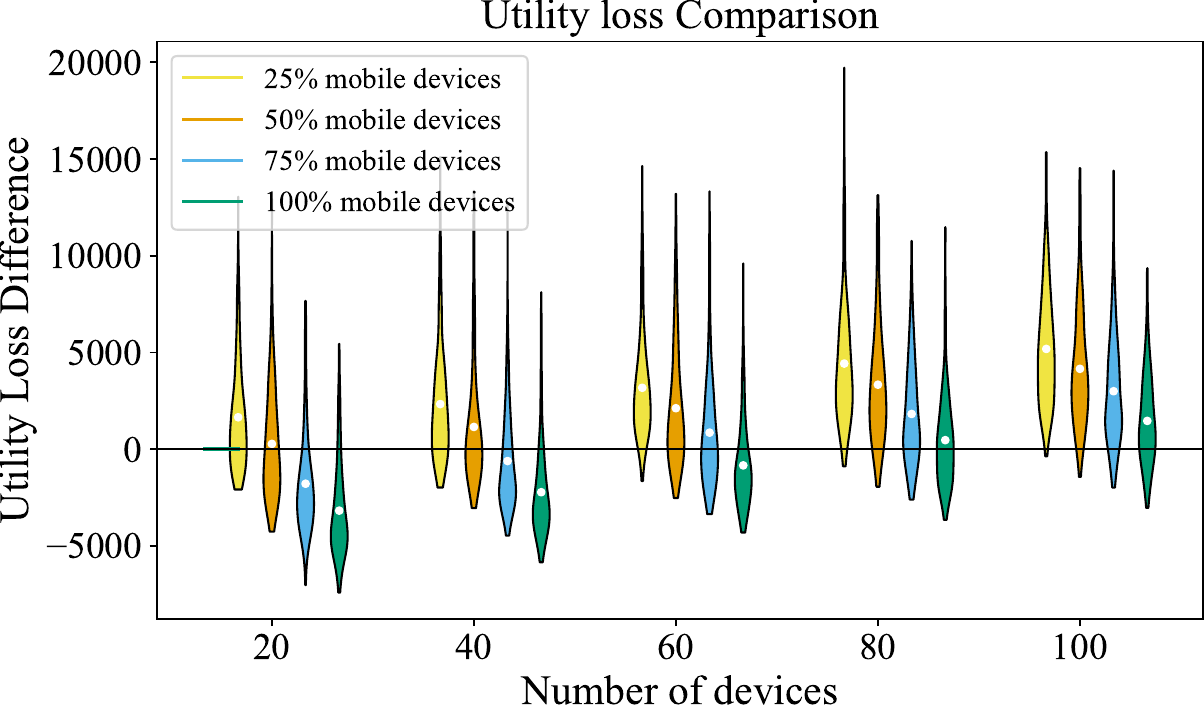}
	\caption{The difference of utility loss achieved with and without mobility}
	\vspace{-2mm}
	\label{fig:heuristic}
\end{figure}

Based on parameters from Table~\ref{tab.simparams}, 50 samples of each aggregator class combination with different device utilizations are generated and simulated. 
The simulation results are shown in Figure~\ref{fig:heuristic}.
Axes of the violin plot represent the utility loss plotted as a function of the number of devices utilizing the aggregator capacity as listed in Table~\ref{tab.simparams}.
Each violin in the plot represents the difference of utility loss difference between device mobility enabled and disabled, \emph{i.e.,} $u_k(t) \mathrm{(mobility)} - u_k(t) \mathrm{(non-mobility)}$.
The former allows devices to move across different clusters while the latter restricts all devices to their original cluster.
The length of the violin indicates the variance while the average value is indicated by the white dot at the center; the sharp peaks represent the outliers.

The observed loss with mobility-enabled was higher than that of mobility-disabled for a lower number of devices and a reversed pattern was observed for a higher number of devices.
For a fixed capacity of the aggregator, a lower number of devices translates to devices with higher demand (\emph{e.g.} electric buses, etc.) while a higher number of devices have devices with lower demand (\emph{e.g.} HVACs, e-scooters, etc.).
Movement may not be effective to reduce the utility loss due to the higher demand of the devices for a smaller number of devices. 
Scheduling a mobile device with high demand (and associated power modes) is difficult at the aggregator as devices within its cluster also have a high demand resulting in long wait times.
Consequently, there is a higher utility loss due to mobility and deadline misses at the devices based on the resulting schedule.

As seen from Figure~\ref{fig:heuristic}, with limited devices, the demand reduces from each of the devices, giving a higher probability of allocating the residual power to a new device.
This can be seen by the reduction in utility loss for mobility enabled in the plot for a higher number of devices. 
The scheduling can be realized practically as electric buses are more difficult to schedule than a low-power EV or electric scooter.
%The violpplot shows two boxes for with and without mobility incorporated into the heuristic. 
%Although there is a range of values spanned by both the , mobility enables devices to move to other aggregators where energy can be obtained in case of unavailability at the original aggregator.
%The higher losses of utility for mobility can be attributed to unknown future loads, \emph{i.e.,} there could be a high load on the target aggregator after a device moves from a different aggregator.
%It is also seen from the plot that there is a higher benefit of mobility when the load on the aggregator is high. 
%This is especially useful during peak times when high load is observed.

\subsection{Comparison with Solver}
Typically, optimization problems are solved with a solver since a "near-optimal" solution can be obtained. 
%In this experiment, we compared the performance of the proposed heuristic with the solutions obtained from the Gurobi solver.
Although a solver can produce an optimal solution in practice, the runtime to complete the simulation is prohibitive. 
In our experiments, the solver runtime was observed to be upwards of $10$ hours.
To prevent the long runtime, a time limit is set to the maximum time taken by the heuristic solution to terminate the simulation to obtain a feasible solution.
%As the solver takes a prohibitive time to complete the simulation, we set a time limit to the maximum time taken by a heuristic sample to stop the simulation.
Additionally, we picked 10 random samples of the 50 samples in the previous experiment to test the performance of the solver.
%With each time slot mapping to 30 minutes, the maximum observed time from the heuristic allows for a pragmatic schedule computation.
For the same time limit, we compare the results of the solver to those from the heuristic and plot the difference in the utility loss \emph{i.e.,} $u^{solver}_k(t) - u^{heuristic}_k(t)$ with mobility enabled.

\begin{figure}[ht]
	\centering  
	\includegraphics[width=0.9\linewidth]{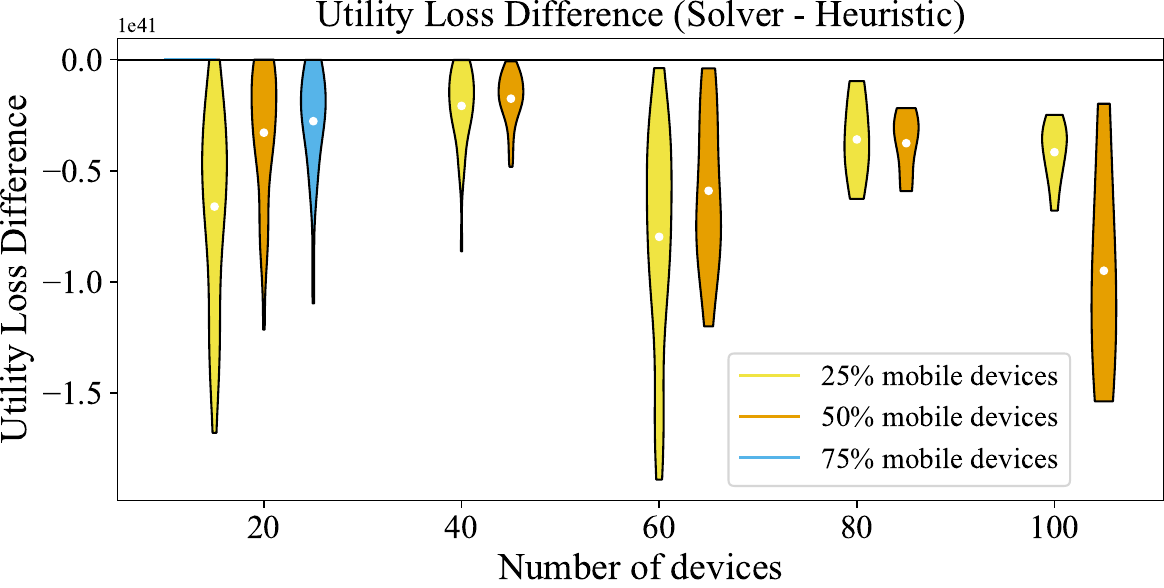}
	\caption{Comparison of solver performance with the heuristic for the same runtime.}
	\vspace{-2mm}
	\label{fig:solver}
\end{figure}

\begin{table}[t]
	\begin{minipage}[b]{1.0\linewidth}
		\begin{center}
			\caption{Number of samples generated for the solver.}
			\vspace{-2mm}
			\label{tab.solverruns}
			\scalebox{0.9}{%
				\begin{tabular}{|l|lll|}
				\hline
				\multirow{2}{*}{No. of devices} & \multicolumn{3}{l|}{No. of Samples} \\ \cline{2-4} 
				& 25\%       & 50\%      & 75\%      \\ \hline
				20                              & 40         & 40        & 5         \\
				40                              & 40         & 13        & 0         \\
				60                              & 40         & 6         & 0         \\
				80                              & 40         & 7         & 0         \\
				100                             & 40         & 9         & 0         \\ \hline
			\end{tabular}

			}
		\end{center}
	\end{minipage}
\end{table}

From Figure~\ref{fig:solver}, we see that even in the best case, the utility loss from the solver is worse off by  $10^{33}$ or more than the heuristic solution.
It can also be seen that the increase in number of devices also increases the losses from the solver. 
%Since the solver needs to compute several stages such as pre-solve, barrier, etc. before the start of optimization, the limited runtime restricts the solver depending on the number of variables. 
Table~\ref{tab.solverruns} lists the number of samples that could produce at least one feasible schedule for the different number of devices along with the proportion of mobile devices among them.
With a lower number of devices and a lower proportion of mobile devices among them, the model is able to produce at least one solution due to lower complexity.
As the proportion of mobile devices increase, the number of solver variables and the resulting losses are exacerbated yielding lesser samples that can produce a feasible schedule.
%Consequently, the samples that could complete execution reduced significantly with higher proportion of mobile devices as seen from Table~\ref{tab.solverruns}.
%This is seen from Table~\ref{tab.solverruns} when 75\% of all devices were mobile, only 20 devices had a few samples that could complete.
%No sample for 100\% mobile devices could complete the run as the time limit expired before finding a solution.

%As seen from the Figure~\ref{fig:solver}, the solver produces a more optimal result against the heuristic at the cost of a prohibitive runtime.
%Similar to the previous section, we compare the utility loss of the system between the solver and heuristic for 50 samples of the chosen aggregator utilization combination.
%It was observed that the heuristic produces roughly the third or fourth best of the solutions provided by the solver.
%This is an acceptable result as the heuristic completes within \_\_\_\_\_ minutes with \_\_\_\_\_ seconds per time slot.
%In a practical scenario, a solution obtained within \_\_\_\_\_ seconds is feasible as a wait time of \_\_\_\_\_ hours for the optimal solution is impractical.

%The impact on the number of aggregators was tested and found that the time increases linearly with increasing aggregators \emph{i.e.,} $\mathcal{O}(l)$.

\begin{table}[t]
	\begin{minipage}[b]{1.0\linewidth}
		\begin{center}
			\caption{Utility loss for EV dataset with different strategies.}
			\vspace{-2mm}
			\label{tab.realworld}
			\scalebox{0.9}{%
	\begin{tabular}{|l|l|l|}
		\hline
		Proposed Heuristic & Earliest Deadline & Highest Power  \\ \hline
	1606.63	&     3939.27              &  3757.18               \\ \hline
	\end{tabular}}
\end{center}
\end{minipage}
\vspace{-5mm}
\end{table}

\subsection{Real-world dataset}
We also simulated the heuristic and the solver on a real-world dataset derived from the EV testbed at Caltech~\cite{3328313}.
The complete charging dataset of the year 2020 was extracted from the testbed data along with data from Table~\ref{tab.simparams} for missing parameters such as power modes, etc.
The arrival times were mapped to a single day (48 time slots) as there was no congestion found with a limited number of vehicles.

The results are shown in Table~\ref{tab.realworld}.
We also tested the dataset with two standard scheduling algorithms: earliest deadline schedules devices with the nearest deadline to the current time slot while highest power schedules devices with the highest demand.
The solution of the proposed heuristic is 59.21\% and 57.23\% better than the earliest deadline and highest power scheduling algorithms, respectively.
The solver could not produce a solution even after 8 hours while the slot time duration ($\tau$) is 30 minutes.
This duration (8+ hours) is impractical for any device arriving with a deadline to know if it can be scheduled, \emph{e.g.,} an EV charing overnight waits without a schedule.
The heuristic produced a solution in one minute with a per slot average time of 1.25 seconds showing the practicality of our proposed solution.
\section{Conclusion}
\label{sec:conclusion}

In this paper, we exploit the device properties such as power modes and mobility to address the issue of load balancing and scheduling.
%Existing demand response solutions in smart grids do not integrate different power modes of the devices as well as their mobility property.
Our model integrates these properties to estimate the utility loss incurred due to delays in receiving the power.
We proposed an online low-complexity heuristic to minimize the utility loss with a practically feasible runtime compared to a solver with similar runtime. 
Our solution also performed better by over 57.23\% on a real-world dataset compared to existing scheduling solutions.
%A real-world testbed data was also used to verify the model and obtain a feasible schedule. 
%As future work, the heuristic can be improved to incorporate more device attributes.
As future work, the heuristic partially integrated with the solver can speed up the solver and improve the utility loss closer to the optimal solution.
%\vspace{-0.5mm}

%{\footnotesize
%\fontsize{8.6}{9.0}\selectfont
\bibliographystyle{IEEEtran}
\bibliography{all}

\end{document}